\documentclass{article}
\usepackage{graphicx}
\usepackage{amsmath}
\usepackage{a4}

\input{tcilatex}

\begin{document}

\title{Response of massive bodies to gravitational waves}
\author{Ludger Hannibal and Jens Warkall \\
Fachbereich Physik, Carl von Ossietzky Universit\"{a}t Oldenburg, Germany}
\maketitle

\begin{abstract}
The response of a massive body to gravitational waves is described on the
microscopic level. The results shed a new light on the commonly used
oscillator model. It is shown that apart from the non-resonant tidal motion
the energy transfer from a gravitational wave to an electromagnetically
coupled body is in general restricted to the surface, whereas gravitational
coupling gives rise to bulk excitation of quadrupole modes, but several
orders of magnitude smaller. A\ microscopic detector making use of the
effect is suggested.
\end{abstract}

\section{Introduction}

Gravitational waves were already considered by Einstein as the wave
solutions of the linearized field equations of gravity. There is indirect
evidence of their existence through systems of binary pulsars that loose
energy in form of gravitational radiation \cite{will:1993}, their direct
experimental measurement presently is one of the most challenging tasks in
gravitational physics. Very sensitive detectors operating at the quantum
limit are needed to detect directly gravitational waves from cosmic events
such as collapsing or colliding star systems. There are basically two
different types of detectors: resonant mass antennas based on the resonant
excitation of quadrupole-type modes of a appropriately chosen massive body,
like the bar detectors conceived by Weber \cite{mtw}, and laser
interferometric devices that detect the direction-dependent variation of the
proper distance between the mirrors of a Michelson interferometer \cite
{saulson:1994}. Detectors of both types are presently under construction 
\cite{list}.

Commonly a resonant mass antenna is described in Riemannian normal
coordinates with respect to its center of mass, the proper frame of
reference (PFR). The detector is analyzed in term of normal modes, idealized
by a spring that couples two masses \cite{mtw}, the resonant energy input is
calculated. The intention of the work presented in this article was twofold:
First, to validate the results obtained from the normal mode model by
microscopic considerations, second to give a complementary description of
the detector in the reference system of the wave, in which the linearized
solutions of the Einstein field equations are computed. Our model is based
on the local properties of a detector in terms of the fundamental binding
forces, electromagnetic and other, which we consider in both the PFR and the
wave system. Analyzing the deviations from the tidal motion, we find that
the energy input from gravitational waves on an electromagnetically coupled
massive body is restricted to the surface of the body, whereas gravitational
coupling leads to true bulk excitation of quadrupole modes. This result does
not contradict the normal mode picture at all, rather it presents a
complementary viewpoint that has eluded the normal mode analysis. The reason
is that though the energy input into any single mode is nonlocal, in the
special case of gravitational waves the superposition of all excited modes
describes a localized excitation. Based on our observations we propose a
new, microscopic type of detector.

\section{Frames of reference}

There exist two frames of reference that can be used for the analysis of
gravitational waves and gravitational wave detectors. The natural system to
study the waves is a perturbed Minkowski system for which the linearized
Einstein field equations are solved. In this reference system the plane wave
is gauged, conventionally a transverse-traceless (TT) gauge is chosen \cite
{mtw}. On the other hand, the use of the PFR system with Riemannian normal
coordinates with respect to the center of mass is the natural system for the
study of a detector. For our purpose to study of the detector on the
microscopic level, both system have advantages and disadvantages. In the PFR
system we have common argument \cite{thorne:1983} that in the metric of the
gravitational wave field leads to variations of the electromagnetic field of
order 
\begin{equation}
\delta A/A\sim \left( L^{2}/\lambda ^{2}\right) h^{TT}  \label{i1}
\end{equation}
where $L$ is the distance from the origin, $\lambda $ and $h^{TT}$ \ the
wave length and amplitude of the gravitational wave, respectively, whereas
the tidal gravitational forces lead to displacements of the constituent
particles of the body that in turn lead to variations of the electromagnetic
field of order 
\begin{equation}
\delta A/A\sim h^{TT}  \label{i2}
\end{equation}
so that the metric effect can be neglected. This means that in the PFR
system the unperturbed solutions of the Maxwell equations can be used. On
the other hand, in the PFR system \ we have to deal with the general problem
of general relativity that local energy densities have no invariant meaning,
so that we cannot easily control the exchange of energy between the wave and
the detector. Here the TT system has the advantage that we can employ a
Hamiltonian description of the constituent particles, and because the system
is in principle a special relativistic one, we have the standard laws of
energy-momentum conservation. Therefore the TT system proves useful for our
considerations. Because considerations in both coordinate system contribute
to the understanding of the detector response, we formulate our results in
both systems.

We first show that in the TT system no energy is transferred to a system of
non-interacting particles;\ in the PFR system this corresponds to the
well-known quadrupole type oscillations. Next we extend the discussion to
electromagnetically coupled systems, where we can use the standard Coulomb
force in the PFR system, whereas we have to solve the metric Maxwell
equations in the TT system. Finally we discuss a gravitationally coupled
system and show that there exists a decisive difference in the response of
the detector which is due to the different nature of the fundamental forces.

\section{Particle Motion}

A point-like test mass with electric charge $e$ can be described in the
presence of electromagnetic fields and gravitation by the Hamiltonian \cite
{hannibal:1991} 
\begin{equation}
H=c\sqrt{\left( p_{\mu }-eA_{\mu }\right) g^{\mu \nu }\left( p_{\nu
}-eA_{\nu }\right) }  \label{1}
\end{equation}
where we use coordinates $x^{\mu },\mu =0,1,2,3,4$ for space-time with
metric $g_{\mu \nu }(x^{\kappa })$ and signature $+---$, $p_{\mu }$ are the
momentum coordinates in the cotangent space, and $A_{\mu }(x^{\kappa })$ is
the electromagnetic four-potential. We also use the notation $\left(
ct,x,y,z\right) $ in an obvious manner. The evolution parameter will be
denoted by $\tau $. The Hamiltonian is conserved, $\partial H/\partial \tau
=0$, it represents the rest mass $m=H/c^{2}$of the particle. The canonical
equations of motion are given by 
\begin{equation}
\dot{x}^{\mu }=\frac{\partial H}{\partial p_{\mu }},\qquad \dot{p}_{\mu }=-%
\frac{\partial H}{\partial x^{\mu }}.  \label{2}
\end{equation}
The constancy of $H$ is equivalent to $\dot{x}^{\mu }g_{\mu \nu }\dot{x}%
^{\nu }=c^{2}$ for any trajectory, thus the evolution parameter is the
proper time.

A gravitational wave propagating in $z$-direction with $+$ and $\times $
polarization modes is described in TT gauge by the metric tensor 
\begin{equation}
g_{\mu \nu }=\left( 
\begin{array}{llll}
1 & 0 & 0 & 0 \\ 
0 & -1+f_{+}(ct-z) & f_{\times }(ct-z) & 0 \\ 
0 & f_{\times }(ct-z) & -1-f_{+}(ct-z) & 0 \\ 
0 & 0 & 0 & -1
\end{array}
\right) =\eta _{\mu \nu }+h_{\mu \nu }  \label{3}
\end{equation}
where $h_{\mu \nu }$ is a small perturbation of the Minkowski metric $\eta
_{\mu \nu }$ \cite{mtw}. This perturbation acts as classical, special
relativistic field, so that the system can be treated within special
relativity, except that the interpretation of the energy must be treated
with care \cite{hannibal:1996}. In the absence of an electromagnetic field
we obtain for this metric a Hamiltonian that leads to four conserved
quantities: $H,p_{x},p_{y}$, and $p_{0}+p_{3}=E/c+p_{z}$ where $E$ is the
energy of the particle in the sense of special relativity. Note that $%
p_{3}=-m\dot{z}$, so that the difference between the energy and the
conventional $z$-momentum is conserved. This is natural, since the
gravitational wave not only carries energy $E_{w}$, but also momentum $P_{w}$
with the relation $E_{w}=c\left| P_{w}\right| $ that holds for all massless
objects in special relativity. The exchange $\Delta E$ of energy between the
wave and a test mass thus is always accompanied with an exchange $\Delta P$
of momentum: 
\begin{equation}
\Delta E=c\Delta P.  \label{4}
\end{equation}
The existence of four conserved quantities now allows us to integrate the
equations of motion completely: 
\begin{eqnarray}
&&
\begin{array}{ll}
\dot{x}^{0}=\frac{1}{m}p_{0} & \dot{x}^{2}=\frac{1}{m}\left(
g^{21}p_{1}+g^{22}p_{2}\right)  \\ 
\dot{x}^{1}=\frac{1}{m}\left( g^{11}p_{1}+g^{12}p_{2}\right)  & \dot{x}^{3}=-%
\frac{1}{m}p_{3} \\ 
p_{1}=const & p_{2}=const
\end{array}
\\
p_{0} &=&\frac{1}{2}\left( p_{0}+p_{3}\right) +\frac{1}{2}\frac{m^{2}}{%
\left( p_{0}+p_{3}\right) }\left( c^{2}-\frac{1}{m^{2}}p_{a}g^{ab}p_{b}%
\right)   \notag \\
p_{3} &=&\frac{1}{2}\left( p_{0}+p_{3}\right) -\frac{1}{2}\frac{m^{2}}{%
\left( p_{0}+p_{3}\right) }\left( c^{2}-\frac{1}{m^{2}}p_{a}g^{ab}p_{b}%
\right)   \label{5}
\end{eqnarray}
where we use the indices $a,b$ for a summation over $1,2$ only. Let us
consider a wave pulse. We denote the initial conditions before the arrival
of the pulse by $\bar{p}_{\mu }$, and have 
\begin{eqnarray}
p_{0}(\tau ) &=&\bar{p}_{0}-\frac{1}{2}\frac{p_{a}h^{ab}p_{b}}{\bar{p}_{0}+%
\bar{p}_{3}}  \notag \\
p_{3}(\tau ) &=&\bar{p}_{3}+\frac{1}{2}\frac{p_{a}h^{ab}p_{b}}{\bar{p}_{0}+%
\bar{p}_{3}}  \label{6}
\end{eqnarray}
where $h^{ab}=g^{ab}-\eta ^{ab}$. Thus after the pulse, where the
perturbation $h$ is zero again, the particle has the same four-momentum as
before, and the only possible effect is a displacement of the straight
trajectory after the pulse from the one before the pulse. If a particle is
initially at rest, it stays at rest in this reference frame. Two particles
that are at rest relative to each other, remain at rest relative to each
other, though the proper distance between them changes with the wave
amplitude. Thus free test particles do not take up energy from a
gravitational wave. When we restrict this statement to a comparison of the
energy before and after a wave pulse or wave train, we are moreover free
from the ambiguity of the energy definition in general relativity.

We now look at this result in the PFR system. In the PFR system particles do
not stay at rest in the wave field, but move under tidal accelerations. In
PFR coordinates $\left( \hat{t},\hat{x},\hat{y},\hat{z}\right) $ the tidal
accelerations produced by the wave are \cite{mtw}: 
\begin{equation}
\frac{d^{2}\hat{x}^{a}}{d\hat{t}^{2}}=-\hat{R}_{a0b0}\hat{x}^{b}=\frac{1}{2}%
\frac{d^{2}h_{ab}}{d\hat{t}^{2}}\hat{x}^{b}  \label{pfr1}
\end{equation}
For a particle that is initially at rest at $\hat{x}^{a(0)}$ before a wave
pulse arrives, the solution is simply given by 
\begin{equation}
\hat{x}^{a}\left( \hat{t}\right) =\hat{x}^{a(0)}+\frac{1}{2}h_{ab}\left( 
\hat{t}\right) \hat{x}^{b(0)}.  \label{pfr2}
\end{equation}
(Note that $\hat{R}_{a0b0}$ and $\hat{R}{}^{a}{}_{0b0}$ differ only by terms
of order $O\left( h^{2}\right) $). This is the well-known quadrupole-like
tidal motion \ of a system of particles around the origin of the frame of
reference. The coordinate distance of the particles varies according to the
changing metric distance between them. We can ascribe a standard kinetic
energy to this motion, but there will be additional contributions to the
conserved energy from the metric. Our analysis in the TT system has shown
that the energy of the wave does not vary in this case. The coordinate
transformation from the TT to the \ PFR system is time-dependent, so that
the energy conservation in the TT\ system translates into a more complicated
conservation law in the PFR system.

Because free particles do not effectively take up energy from a
gravitational wave, we have to take the coupling between particles into
account in order to describe the response of a detector. The coupling is
basically of electromagnetic nature in small massive bodies, but may also be
of gravitational nature in large bodies. Because the effects of the
gravitational wave are so small, massive detectors must be cooled to zero
temperature as near as possible. Thus the ground state of a body where the
constituent atoms or ions are at rest relative to each other except for
quantum effects\ will serve as an appropriate model. Rotational motion with
respect to the TT frame of reference must be taken into account. Internal
motion, thermal or otherwise, of the atoms leads to forces of order $m\Delta
v\partial h/\partial t$; we do not consider this kind of phonon-graviton
interaction in this article, because it is the idea of the Weber detector to
excite the quadrupole modes, not to enhance already excited modes.

\section{Electromagnetic Field}

We first look at the Coulomb potential generated by a charge that resides in
the field of the wave. In the PFR system we have the above cite argument
that the metric perturbations of the electromagnetic potentials can be
ignored. Thus, for the description of an electromagnetically bound solid
body, we have to take only the Coulomb forces into account, and can ignore
contributions from magnetic fields and electromagnetic radiation. In the PFR
system the coordinate distance agrees with the metric distance up to order $%
O(h)$. Therefore we can state in an invariant manner, that the Coulomb
potential depends only on the metric distance \ of the particles. We now
show how this translates into the TT system.

We assume that the wave field is slowly varying and the velocity of the
charge relative to the source of the fields is so small that magnetic fields
can be ignored. The corresponding equation to solve for the electromagnetic
potential generated by a charge $q_{s}$ at rest at $x=y=z=0$ is 
\begin{equation}
\partial _{\mu }\sqrt{-g}g^{\mu \nu }\partial _{\nu }A^{0}=\frac{q_{s}}{%
\varepsilon _{0}}\delta ^{3}(x,y,z).  \label{7}
\end{equation}
Our considerations in the PFR system give rise to the following ansatz: 
\begin{equation}
A_{s}^{0}(\vec{r},ct-z)=\frac{q_{s}}{4\pi \varepsilon _{0}r},\quad \vec{r}%
=\left( x^{i}\right) _{i=1,2,3},\quad r^{2}=-x^{i}g_{ij}\left( ct-z\right)
x^{j}\quad .  \label{8}
\end{equation}
where the Euclidean coordinate distance from the source is replaced by the
metric distance. We verify that equation (\ref{7}) is satisfied to order $%
O(h)$. For simplicity only we consider a wave with the $+$-mode only where 
\begin{equation}
r^{2}=x^{2}(1-f_{+}\left( ct-z\right) )+y^{2}\left( 1+f_{+}\left(
ct-z\right) \right) +z^{2}.  \label{pfr3}
\end{equation}
Using $\sqrt{-g}=1+O\left( h^{2}\right) $, leaving out terms of order $h^{2}$
and higher, we obtain : 
\begin{eqnarray}
\partial _{\mu }\sqrt{-g}g^{\mu \nu }\partial _{\nu }\frac{1}{r} &\cong
&\left( c^{-2}\partial _{t}^{2}-\frac{1}{1-f_{+}}\partial _{x}^{2}-\frac{1}{%
1+f_{+}}\partial _{y}^{2}-\partial _{z}^{2}\right) \frac{1}{r}  \notag \\
&=&c^{-1}\partial _{t}\left( \frac{x^{2}-y^{2}}{2r^{3}}f_{+}^{\prime
}\right) +\partial _{x}\frac{x}{r^{3}}+\partial _{y}\frac{y}{r^{3}}-\partial
_{z}\left( -\frac{z}{r^{3}}-\frac{x^{2}-y^{2}}{2r^{3}}f_{+}^{\prime }\right) 
\notag \\
&=&\nabla \frac{\vec{r}}{r^{3}}+\frac{3\left( x^{2}-y^{2}\right) ^{2}}{4r^{5}%
}\left( f_{+}^{\prime }\right) ^{2}+\frac{x^{2}-y^{2}}{2r^{3}}f_{+}^{\prime
\prime }-\frac{x^{2}-y^{2}}{2r^{3}}f_{+}^{\prime \prime }+f_{+}^{\prime
}\partial _{z}\frac{y^{2}-x^{2}}{2r^{3}}  \notag \\
&=&4\pi \delta \left( r\right) +\frac{3\left( y^{2}-x^{2}\right) ^{2}}{4r^{5}%
}\left( f_{+}^{\prime }\right) ^{2}+f_{+}^{\prime }\partial _{z}\frac{%
y^{2}-x^{2}}{2r^{3}}  \label{pfr4}
\end{eqnarray}
Now the first of these two remaining terms is of order $O\left( h^{2}\right) 
$ and can thus be ignored, the second is proportional to the spatial
derivative of the wave field, which can be ignored for detectors that are
small against the wave length \cite{mtw}. Thus the ansatz (\ref{8}) solves
equation (\ref{7}) except for terms of order $O\left( h^{2}\right) $ or
proportional to $\partial _{z}h$. In the same sense the Lorentz gauge holds.
The result thus agrees with that in the PFR system: The Coulomb potential
depends only on the metric distance.

In the following we assume that this principle may also by extended to
other, phenomenological potentials, because the time scale set by the
gravitational waves is by far larger than that of the induced changes of all
other fundamental interactions. So in the TT system the Coulomb force
between two charged particles varies in phase with the gravitational wave.
It is not hard to see that the local energy density of the electromagnetic
field, though time-dependent, is only relocated, so that the integrated
energy density does not change to first order in $h$, implying that
radiation effects are at most of second order in $h$, in agreement with the
situation in the PFR system. On the other hand accelerated, moving charges
produce magnetic fields and electromagnetic radiation, but both effects can
be ignored for the analysis of weak mode excitation in a solid body.

\section{Many particles}

The main point of this work is to show that a detailed microscopic,
many-particle model of the detector exhibits two aspects that elude the
spring model and the normal mode analysis. The first aspect is that the
input of energy from the wave to the detector occurs only at the surface of
the detector. This effect can already be modelled with a linear chain , as
done in the next subsection. The second aspect is the dependence on the
nature of the fundamental forces that stabilize a body: electromagnetic and
gravitational binding forces lead to different mode structure of the
responde, as shown subsequently.

\subsection{Linear chain model}

In the PFR system $N$ classical particles with coordinates $\vec{r}^{\left(
s\right) }=\left( \hat{x},\hat{y},\hat{z}\right) ,\alpha =1,...,N$ obey the
equations of motion 
\begin{equation}
m_{i}\frac{d^{2}}{d\hat{t}^{2}}\vec{r}^{\left( s\right) }=\sum_{s^{\prime
}\neq s}\vec{F}_{s^{\prime }s}\left( \vec{r}^{\left( s^{\prime }\right) },%
\vec{r}^{\left( s\right) }\right) +\frac{m_{i}}{2}\frac{d^{2}h}{d\hat{t}^{2}}%
\cdot \vec{r}^{\left( s\right) }  \label{pfr5}
\end{equation}
where for simplicity the dot denotes the multiplication of the $3\times 3$
matrix $h_{ij}$ with the vector $\vec{r}^{\left( s\right) }$, and $%
F_{s^{\prime }s}$ represents the fundamental forces between two particles.
Let us consider the special model of a linear chain in $\hat{x}$-direction
with $N=2n$ equal particles numbered by $s^{\prime },s=-n...n,$ coupled by
nearest-neighbor forces 
\begin{equation}
F_{s^{\prime }s}\left( \hat{x}^{\left( s^{\prime }\right) },\hat{x}^{\left(
s\right) }\right) =-\omega _{0}^{2}\left( \hat{x}^{\left( s^{\prime }\right)
}-\hat{x}^{\left( s\right) }-l_{0}\right) \text{ for }\left| s^{\prime
}-s\right| =1  \label{pfr6}
\end{equation}
and a wave with $+$-mode only, so that the equations of motion are 
\begin{equation}
\frac{d^{2}}{d\hat{t}^{2}}\hat{x}^{\left( s\right) }=-k\sum_{\beta =\alpha
\pm 1}\left( \hat{x}^{\left( s\right) }-\hat{x}^{\left( s^{\prime }\right)
}\pm l_{0}\right) +\frac{1}{2}\frac{d^{2}f_{+}}{d\hat{t}^{2}}\hat{x}^{\left(
s\right) }  \label{pfr7}
\end{equation}
where $k=\omega _{0}^{2}/m$. Without the tidal acceleration, the equilibrium
positions are $\hat{X}^{\left( s\right) }=sl_{0}$, relative to the center of
mass that is identical with the origin of the coordinate system. We are now
interested in the deviation from the tidal motion (\ref{prf2}) because the
tidal motion itself will be in general too small to be observed itself. We
set 
\begin{equation}
\xi ^{\left( s\right) }=\hat{x}^{\left( s\right) }-\hat{X}^{\left( s\right)
}\left( 1+\frac{f_{+}}{2}\right)   \label{pfr8}
\end{equation}
leading to 
\begin{eqnarray}
\ddot{\xi}^{\left( s\right) } &=&-k\sum_{s^{\prime }=s\pm 1}\left( \xi
^{\left( s\right) }-\xi ^{\left( s^{\prime }\right) }-\left( \hat{X}^{\left(
s\right) }-\hat{X}^{\left( s^{\prime }\right) }\right) \left( 1+\frac{f_{+}}{%
2}\right) \pm l_{0}\right) +\frac{1}{2}\frac{d^{2}f_{+}}{d\hat{t}^{2}}\left( 
\hat{x}^{\left( s\right) }-\hat{X}^{\left( s\right) }\right)   \notag \\
&=&-k\sum_{s^{\prime }=s\pm 1}\left( \xi ^{\left( s\right) }-\xi ^{\left(
s^{\prime }\right) }\pm l_{0}\frac{f_{+}}{2}\right)   \label{pfr9}
\end{eqnarray}
where we have omitted the term $\frac{1}{2}\frac{d^{2}f_{+}}{d\hat{t}^{2}}%
\left( \hat{x}^{\left( s\right) }-\hat{X}^{\left( s\right) }\right) $
because we consider deviations from the tidal motion that are proportional
to $f_{+}$ itself, as induced by the wave, and thus this term is of order $%
O\left( h^{2}\right) $. Now the following happens: the terms $l_{0}\frac{%
f_{+}}{2}$ in (\ref{pfr9}) cancel for all particles that have two neighbors,
only for the particles \ at the ends of the chain they do not. The reason is
that under the tidal motion the distances between pair of neighboring
particles remains constant, thus the induced additional forces from the left
and right cancel; the pattern of the tidal acceleration comes very close to
a null mode. So we can rewrite 
\begin{equation}
\ddot{\xi}^{\left( s\right) }=-k\sum_{s^{\prime }=s\pm 1}\left( \xi ^{\left(
s\right) }-\xi ^{\left( s\right) }\right) -kl_{0}\frac{f_{+}}{2}\left(
\delta _{s,n}-\delta _{s,-n}\right) .  \label{pfr10}
\end{equation}
These equations have a clear interpretation: The deviation from the tidal
motion is driven by an effective force that applies to the ends of the chain
only. Or, we can state that it suffices to apply additional forces $\mp
kl_{0}f_{+}/2$ to the ends of the chain in order to suppress any deviations
from the tidal motion. In this case the tension of the chain varies
uniformly along with the wave strength, but no work is done at all against
the internal coupling forces.

Regarding the energy we have to be careful to use standard expressions,
because there are metric contributions, as we have seen above, but we can
certainly state that apart from the tidal motion energy is transferred to
the chain only locally at the ends. For resonance cross terms in the kinetic
energy between the tidal motion and the deviations from it play no role, so
that $\sum m\dot{\xi}^{2}/2$ describes the kinetic energy of interest.
Nevertheless this local picture does not contradict the normal-mode analysis
in any way. We still have the possibility to decompose the perturbing force
into normal modes and derive equations for the driving of these modes in the
standard way. The fundamental mode of the chain is driven strongest, but all
other symmetric modes are also excited. It is the superposition of all these
modes that gives rise to the local excitation of the chain. Only refining
the spring model to a linear chain model could exhibit this property.

Because we see from (\ref{pfr10}) that the effective force depends on the
microscopic property of the chain in form of the equilibrium distance $l_{0}$%
, it is necessary to analyze a detector on the basis of a more realistic
microscopic model that yield information on the strength of the driving
forces. It turns out that this is also necessary because the difference
between gravitational and electromagnetic coupling can only then be shown.

\subsection{Microscopic forces in the PFR system}

For a more realistic model we can use eq. (\ref{pfr5}) with the fundamental
Coulomb, gravitational, and repulsive short-range forces inserted. We assume
that all these forces depend only on the difference vectors $\vec{r}^{\left(
\beta \right) }-\vec{r}^{\left( \alpha \right) }$ and stable equilibrium
positions $\vec{R}^{\left( \alpha \right) }$ exists. Again we look at the
deviation from the tidal motion induced by these forces. The transformation 
\begin{equation}
\vec{r}^{\left( \alpha \right) }=\left( 1+\frac{1}{2}h\right) \cdot \vec{\rho%
}^{\left( \alpha \right) }  \label{pfr11}
\end{equation}
now leads, assuming small deviations from equilibrium, 
\begin{equation}
\vec{\rho}^{\left( \alpha \right) }=\vec{R}^{\left( \alpha \right) }+O\left(
h\right) ,  \label{pfr12}
\end{equation}
to 
\begin{eqnarray}
\frac{d^{2}}{d\hat{t}^{2}}\vec{\rho}^{\left( \alpha \right) } &=&\sum_{\beta
\neq \alpha }\vec{F}_{\beta \alpha }\left( \left( 1+\frac{1}{2}h\right)
\cdot \left( \vec{\rho}^{\left( \beta \right) }-\vec{\rho}^{\left( \alpha
\right) }\right) \right) +O\left( h^{2}\right)  \label{pfr13} \\
&=&\sum_{\beta \neq \alpha }\left[ \vec{F}_{\beta \alpha }\left( \vec{\rho}%
^{\left( \beta \right) }-\vec{\rho}^{\left( \alpha \right) }\right) +\frac{1%
}{2}\left( h\cdot \nabla \right) \vec{F}_{\beta \alpha }\left( \vec{\rho}%
^{\left( \beta \right) }-\vec{\rho}^{\left( \alpha \right) }\right) \right]
+O\left( h^{2}\right)  \notag
\end{eqnarray}
We will see that this equation for the deviation is identical with that
derived for the motion in the TT system. The reason is that the
transformation (\ref{prf11}) basically agrees with the coordinate
transformation from the PFR to the TT\ system. For the distance we have 
\begin{equation*}
\vec{r}^{i}\delta _{ij}\vec{r}^{j}=\vec{\rho}^{i}\left( \delta _{ik}+\frac{1%
}{2}h_{ik}\right) \left( \delta _{kj}+\frac{1}{2}h_{kj}\right) \vec{\rho}%
^{j}=-\vec{\rho}^{i}g_{ij}^{TT}\vec{\rho}^{j}+O\left( h^{2}\right)
\end{equation*}
and thus all potential forces acting on the deviation of the tidal motion in
the PFR system are identical with that in the TT system.

\subsection{Hamiltonian analysis in the TT system}

We make a potential approximation for the many-particle Hamiltonian, since
in a fully relativistic approach we had to include necessarily the dynamics
of the electromagnetic field in order to preserve energy-momentum
conservation. Thus we write the total Hamiltonian for many particles with
coordinates $\left( x^{(s)\mu },p_{\mu }^{(s)}\right) $, $s=1,2,...$ as 
\begin{equation}
H=\sum_{s}H^{(s)}  \label{9}
\end{equation}
where 
\begin{equation}
H^{(s)}=-\frac{1}{2m_{1}}p_{i}^{(s)}g^{ij}\left( ct-z^{(s)}\right)
p_{j}^{(s)}+\sum_{s\prime \neq s}\frac{1}{2}V_{ss\prime }(\vec{r}^{(s)}-\vec{%
r}^{(s\prime )},ct-z^{(s\prime )})  \label{10}
\end{equation}
is the contribution from a single particle. $V_{ss\prime }$ is the total
potential generated by the particle $s^{\prime }$, acting on particle $s$.
As it should be, to each particle only half the potential energy is
attributed. In the electromagnetic case, the other half, as well as the
infinite self-energy is subtracted with the contribution from the
electromagnetic field energy \cite{hannibal:1996}. We assume that the
potential is a central potential depending only on the metric distance: 
\begin{equation*}
V_{ss\prime }(\vec{r}^{(s)}-\vec{r}^{(s\prime )},ct-z^{(s\prime
)})=V_{ss\prime }^{0}(\sqrt{-\left( \vec{r}^{(s)}-\vec{r}^{(s\prime
)}\right) ^{i}g_{ij}\left( ct-z^{(s\prime )}\right) \left( \vec{r}^{(s)}-%
\vec{r}^{(s\prime )}\right) ^{j}}),
\end{equation*}
as derived for the Coulomb potential.

The evolution parameter is the time $t$, common to all particles. The
Hamiltonian (\ref{9}) conserves the difference between the total energy and
the center of mass $z$-momentum: 
\begin{equation}
\frac{d}{dt}\left( H+c\sum_{s}p_{3}^{(s)}\right) =0.  \label{11}
\end{equation}
But since we have less conservation laws than coordinates, energy and
momentum transfer from the wave to the particle system has become possible.
The change of the total energy is calculated using the equations of motion 
\begin{eqnarray}
\frac{d}{dt}p_{z}^{(s)} &=&-\frac{\partial H}{\partial z^{(s)}}  \notag \\
\qquad &=&-\frac{1}{2m_{s}}p_{i}^{(s)}h^{ij\prime }\left( ct-z^{(s)}\right)
p_{j}^{(s)}  \notag \\
&&-\frac{1}{2}\frac{\partial }{\partial z^{(s)}}\left( \sum_{s\prime \neq
s}V_{s\prime s}(\vec{r}^{(s)}-\vec{r}^{(s\prime
)},ct-z^{(s)})+\sum_{s^{\prime }\neq s}V_{ss\prime }(\vec{r}^{(s\prime )}-%
\vec{r}^{(s)},ct-z^{(s\prime )})\right)  \notag \\
&&  \label{12}
\end{eqnarray}
that lead us to 
\begin{eqnarray}
\frac{d}{dt}\sum_{s}p_{z}^{(s)} &=&-\sum_{s}\frac{1}{2m_{s}}%
p_{a}^{(s)}h^{ab\prime }\left( ct-z^{(s)}\right) p_{b}^{(s)}  \notag \\
&&+\frac{1}{2}\sum_{s,s^{\prime }\neq s}\partial _{2}V_{s\prime s}(\vec{r}%
^{(s)}-\vec{r}^{(s^{\prime })},ct-z^{(s)})  \label{13}
\end{eqnarray}
where $h^{ij\prime }$ denotes the derivative, $\partial _{2}V_{s\prime s}$
the partial derivative with respect to the second argument only. The
derivatives of $V_{s\prime s}$ with respect to the first argument cancel in
the sum because of the dependence on $\vec{r}^{(s)}-\vec{r}^{(s^{\prime })}$
only. To first order in the perturbation we can approximate 
\begin{equation}
\partial _{2}V_{s\prime s}(\vec{r}^{(s)}-\vec{r}^{(s^{\prime
})},ct-z^{(s)})\simeq -V_{s\prime s}^{0\prime }\left( r_{ss\prime }\right) 
\frac{x_{ss\prime }^{a}x_{ss\prime }^{b}}{2r_{ss\prime }}h_{ab}^{\prime
}\left( ct-z^{(s)}\right)  \label{14}
\end{equation}
where we use $x_{ss\prime }^{a}=\left( \vec{r}^{(s)}-\vec{r}^{(s\prime
)}\right) ^{a}$ for short, and $r_{ss\prime }$ is the unperturbed Euclidean
distance. Thus we have (with $h^{ab}=-h_{ab}$ in lowest order) 
\begin{equation}
\frac{d}{dt}\sum_{s}p_{z}^{(s)}=\sum_{s}h_{ab}^{\prime }\left(
ct-z^{(s)}\right) \left[ \frac{1}{2m_{s}}p_{a}^{(s)}p_{b}^{(s)}-\sum_{s^{%
\prime }\neq s}V_{s\prime s}^{0\prime }\left( r_{ss\prime }\right) \frac{%
x_{ss\prime }^{a}x_{ss\prime }^{b}}{2r_{ss\prime }}\right] .  \label{15}
\end{equation}
For a pulse of duration $T$ that interacts with the particle system the
change of the center of mass $z$-momentum is given by 
\begin{equation}
\Delta P_{z}=\sum_{s}\int_{t_{0}}^{t_{0}+T}h_{ab}^{\prime }\left(
ct-z^{(s)}\right) q_{ab}^{(s)}(t)dt  \label{16}
\end{equation}
where $q_{ab}^{(s)}$ is the microscopic contribution from a single particle: 
\begin{equation}
q_{ab}^{(s)}(t)=\frac{1}{2m_{s}}p_{a}^{(s)}p_{b}^{(s)}-\sum_{s\prime \neq
s}V_{s\prime s}^{0\prime }\left( r_{ss\prime }\right) \frac{x_{ss\prime
}^{a}x_{ss\prime }^{b}}{2r_{ss\prime }}.  \label{17}
\end{equation}
This contribution is quadrupole-like, but it is weighted with derivative of
the potential. Then the sum or the integral over all particles leads in the
case of the Coulomb potential to alternating sums where contributions
cancel, similar to the summation leading to the Madelung constant. On the
other hand, if the potential is purely attractive, as for gravitation, this
effect does not occur, giving rise to a completely different response, as
well will see.

For a small detector we can assume that the wave field is constant over the
particle system represented by the center-of-mass coordinate $z(t)$, so we
can change the integration variable to $\tau =t-z(t)/c$ and obtain 
\begin{equation}
\Delta P_{z}=\int_{-\infty }^{+\infty }d\tau h_{ab}^{\prime }\left( c\tau
\right) \sum_{s}\frac{q_{ab}^{(s)}(t(\tau ))}{1-\dot{z}(t(\tau ))/c}%
=\int_{-\infty }^{+\infty }d\tau h_{ab}^{\prime }\left( c\tau \right)
Q_{ab}(\tau )  \label{18}
\end{equation}
with 
\begin{equation}
Q_{ab}(\tau )=\sum_{s}\frac{q_{ab}^{(s)}(t(\tau ))}{1-\dot{z}(t(\tau ))/c}.
\end{equation}
For a wave pulse or wave train with duration $T$, such that $h_{ab}\left(
0\right) =h_{ab}\left( cT\right) =0$ we use partial integration to write (%
\ref{18}) as 
\begin{equation}
\Delta E=c\Delta P_{z}=-\int_{-\infty }^{+\infty }d\tau h_{ab}\left( c\tau
\right) Q_{ab}^{\prime }(\tau )  \label{19}
\end{equation}
Thus the response of the detector to a gravitational wave pulse travelling
in $z$-direction is described by the time-dependence of the microscopic
function $Q_{ab}$. Alternatively, we may use Fourier transform to arrive at 
\begin{equation}
\Delta E=\int_{-\infty }^{+\infty }d\nu \hat{h}_{ab}^{\ast }(\nu )i\nu \hat{Q%
}_{ab}(\nu )  \label{20}
\end{equation}
with the spectral decompositions of the wave: 
\begin{equation}
\hat{h}_{ab}(\nu )=\int h_{ab}(c\tau )e^{-2\pi i\nu \tau }d\tau ,  \label{21}
\end{equation}
and the particle system: 
\begin{equation}
\hat{Q}_{ab}(\nu )=\sum_{s}\int \frac{q_{ab}^{(s)}(t(\tau ))}{1-\dot{z}%
(t(\tau ))/c}e^{-2\pi i\nu \tau }d\tau .  \label{22}
\end{equation}
Thus $\hat{Q}_{ab}$ represents the effective cross section of the particle
system on the microscopic level.

\section{Force distribution in a massive body}

We first consider electromagnetically \ coupled bodies, as ion crystals and
metals, where the gravitational forces between the constituents play no
role. Ion crystals have the advantage that all charges can be considered to
be point-like and located on lattice points. Thus only discrete sums have to
be evaluated. As we have seen, the forces driving the deviations from the
tidal motion in the PFR system are identical to the forces driving the whole
motion in the TT system as long as we consider only motions of order $%
O\left( h\right) $.

\subsection{\protect\bigskip Ion crystal}

We first consider a gravitational wave propagating in $z$-direction incident
on a lattice of ions. In order for the system to possess a stable ground
state, we have to include not only the Coulomb potential, but also some
(short-ranged) repulsive potential. The Coulomb force exerted by particle $%
s\prime $ on particle $s$ is given, to first order in $h$, by 
\begin{equation}
F_{C,\;ss\prime }^{a}=\frac{q_{s}q_{s\prime }}{4\pi \;\varepsilon _{0}}%
\;\left( \frac{x_{ss\prime }^{a}}{r_{ss\prime }^{3}}-h_{ab}\frac{x_{ss\prime
}^{b}}{r_{ss\prime }^{3}}+\frac{3}{2}h_{bc}\frac{x_{ss\prime
}^{b}x_{ss\prime }^{c}x_{ss\prime }^{a}}{r_{ss\prime }^{5}}\right) ,
\label{24}
\end{equation}
for the $x$- and $y$- directions, the force in $z$-direction additionally
involves the derivative of $h$, which is not of interest here. In the case
of the Born-Meyer potential the repulsive potential is of exponential form 
\begin{equation}
V_{BM}(r_{ss\prime })=A_{ss\prime }e^{-\beta r_{ss\prime }}  \label{25}
\end{equation}
with couplings $A_{ss\prime }$ that depend on the charges. Though it is
necessary to include this potential, its precise form is not crucial for our
further considerations. We obtain an additional force from this potential,
up to first order in $h$ given by 
\begin{eqnarray}
F_{BM,ss\prime }^{a} &=&-\frac{\partial V_{BM,}}{\partial x^{a\;(s)}}, 
\notag \\
&=&A_{ss\prime }\beta \frac{e^{-\beta r_{ss\prime }}}{r_{ss\prime }}%
\;x_{ss\prime }^{a}+\frac{1}{2}\;A_{ss\prime }\beta \;\frac{e^{-\beta
r_{ss\prime }}}{r_{ss\prime }}\;\left( \frac{1}{r_{ss\prime }^{2}}+\beta 
\frac{1}{r_{ss\prime }}\right) h_{bc}\;x_{ss\prime }^{b}\;x_{ss\prime
}^{c}\;x_{ss\prime }^{a}  \notag \\
&&-A_{ss\prime }\beta \;\frac{e^{-\beta r_{ss\prime }}}{r_{ss\prime }}%
\;h_{ab}\;x_{ss\prime }^{b}.  \label{26}
\end{eqnarray}
The total force on ion $s$ is given by 
\begin{eqnarray}
F_{s}^{a} &=&\sum_{s\prime \neq s}\left( \frac{q_{s}q_{s\prime }}{4\pi
\varepsilon _{0}}\;\frac{1}{r_{ss\prime }^{3}}+A_{ss^{\prime }}\beta \frac{%
e^{-\beta r_{ss\prime }}}{r_{ss\prime }}\right) \;x_{ss\prime }^{a}  \notag
\\
&&-h_{ab}\;\sum_{s\prime \neq s}\left( \frac{q_{s}q_{s\prime }}{4\pi
\varepsilon _{0}}\;\frac{1}{r_{ss\prime }^{3}}+A_{ss\prime }\beta \frac{%
e^{-\beta r_{ss\prime }}}{r_{ss\prime }}\right) \;x_{ss\prime }^{b}  \notag
\\
&&+h_{bc}\;\sum_{s\prime \neq s}\left( \frac{3q_{s}q_{s\prime }}{8\pi
\varepsilon _{0}}\;\frac{1}{r_{ss\prime }^{5}}+\frac{1}{2}A_{ss\prime }\beta 
\frac{e^{-\beta r_{ss\prime }}}{r_{ss\prime }}\left( \frac{1}{r_{ss\prime
}^{2}}+\frac{\beta }{r_{ss\prime }}\right) \right) \;x_{ss\prime
}^{b}\;x_{ss\prime }^{c}\;x_{ss\prime }^{a}  \notag \\
&&  \label{27}
\end{eqnarray}
Assuming that the ion chain is in its unperturbed equilibrium state, both
the first and second terms vanish because the vanishing of the first term
defines the equilibrium in absence of a gravitational wave, and the second
term is just the first multiplied by \ the matrix \ $h_{ab}$. Thus the
gravitational wave gives rise to a perturbation of the equilibrium state
induce by the force 
\begin{equation}
\Delta F_{s}^{a}=\left[ h_{bc}\;\sum_{s\prime \neq s}\left( \frac{%
3q_{s}q_{s\prime }}{8\pi \varepsilon _{0}}\;\frac{1}{r_{ss\prime }^{5}}+%
\frac{1}{2}A_{ss\prime }\beta \frac{e^{-\beta r_{ss\prime }}}{r_{ss\prime }}%
\left( \frac{1}{r_{ss\prime }^{2}}+\frac{\beta }{r_{ss\prime }}\right)
\right) \;x_{ss\prime }^{b}\;x_{ss\prime }^{c}\right] \;x_{ss\prime }^{a}
\label{28}
\end{equation}
Once the system is driven out of the equilibrium state, we still can ignore
the second term in (\ref{27}), because for deviations $\Delta x_{ss\prime
}^{a}\sim h$ from equilibrium this term is only of order $h^{2}$. The first
term then describes phonon-graviton interaction.

The sum over all other ion in (\ref{28}) now depends on the dimension of the
lattice. The sum over the short-ranged part converges even for an infinite
lattice, so that its contribution is always limited. The sum over the
Coulomb part is an alternating sum. In one dimension, this sum is always
bounded by the first term that is not canceled by a contribution from a
symmetric neighbor, thus the force has a maximum on the endpoints and
decreases proportional to $R^{-2}$ with the distance $R$ from the endpoints.
In two dimensions we find that the force distribution exhibits the correct
quadrupole structure \cite{diplom}. Therefore acoustic modes are excited.
This result is obtained only when the short-range potential is included. The
Coulomb forces alone gives rise to forces that additionally alternate in
direction from ion to ion, so that we would have arrived at the wrong
conclusion that optical modes are excited.

The forces decay rapidly away from the boundary and do not follow the linear
law of the tidal accelerations. Thus, the deviations from the tidal motion
excite a broad spectrum of modes, of which few are resonantly driven. We now
present a general argument that this pertains to the relevant case of three
dimensions.

\subsection{General case}

Since the temperature of the body must be low in order to achieve the
desired sensitivity of a detector, we assume that the mass is a perfect
crystal, either an ion crystal with discrete charges locate on some lattice,
or a metal with ions on some lattice and the electron gas in between. The
body is decomposed into a finite number of elementary cells $V_{i}$ with
their charge centers at $\vec{r}_{i}$, that are \ (i) electrically neutral, 
\begin{equation}
\int_{V_{i}}d^{D}r\rho (\vec{r})=0  \label{29}
\end{equation}
and (ii) do not possess an electric dipole moment, 
\begin{equation}
\int_{V_{i}}d^{D}r\left( \vec{r}-\vec{r}_{i}\right) \rho (\vec{r})=0.
\label{30}
\end{equation}
$D$ is the spatial dimension of the lattice. We now consider a charge
element $q^{\prime }=\rho (\vec{r}^{\prime })dV$ located at $\vec{r}^{\prime
}$ and the perturbational force exerted on it by the elementary cell $V$
located at $\vec{r}_{i}=0$, according to (\ref{28}). \ For $\left| \vec{r}%
^{\prime }\right| \gg 1/\beta $ we can ignore the short-range potential and
have 
\begin{equation}
F_{s}^{a}\left( \vec{r}^{\prime }\right) =q^{\prime }\int_{V}d^{D}r\,\;\frac{%
3\rho (\vec{r})}{8\pi \varepsilon _{0}}\;\frac{\left( \vec{r}^{\prime }-\vec{%
r}\right) ^{a}}{\left| \vec{r}^{\prime }-\vec{r}\right| ^{5}}h_{bc}\;\left( 
\vec{r}^{\prime }-\vec{r}\right) ^{b}\left( \vec{r}^{\prime }-\vec{r}\right)
^{c}.  \label{31}
\end{equation}
We expand the integrand into powers of $\vec{r}=($ $x,y,z)$ up to second
order. Then the integrals of the first and second order vanish due to
conditions (\ref{29}) and (\ref{30}), respectively. As an example, if $h$ is
of $+$-polarization, the forces in two dimensions are given by 
\begin{eqnarray}
F_{q\prime ,V}^{x}\left( \vec{r}^{\prime }\right)  &=&q^{\prime }h_{+}\left[ 
\frac{x^{\prime }\left( 6x^{\prime 2}-9y^{\prime 2}\right) }{2r^{\prime 7}}%
I_{1}+\frac{x\prime \left( -3x^{\prime 2}+12y^{\prime 2}\right) }{2r^{\prime
7}}I_{2}+\frac{y^{\prime }\left( 12x^{\prime 2}-3y^{\prime 2}\right) }{%
r^{\prime 7}}I_{3}\right]   \notag \\
&&  \label{32} \\
F_{q\prime ,V}^{y}\left( \vec{r}^{\prime }\right)  &=&q^{\prime }h_{+}\left[ 
\frac{y^{\prime }\left( 12x^{\prime 2}-3y^{\prime 2}\right) }{2r^{\prime 7}}%
I_{1}+\frac{y^{\prime }\left( -9x^{\prime 2}+6y^{\prime 2}\right) }{%
2r^{\prime 7}}I_{2}+\frac{x^{\prime }\left( -3x^{\prime 2}+12y^{\prime
2}\right) }{r^{\prime 7}}I_{3}\right]   \notag \\
&&  \label{33}
\end{eqnarray}
where the integrals 
\begin{equation}
I_{1}=\int_{V}d^{D}r\,x^{2}\;\frac{3\rho (\vec{r})}{8\pi \varepsilon _{0}}%
,\quad I_{2}=\int_{V}d^{D}r\,y^{2}\;\frac{3\rho (\vec{r})}{8\pi \varepsilon
_{0}},\quad I_{3}=\int_{V}d^{D}r\,xy\;\frac{3\rho (\vec{r})}{8\pi
\varepsilon _{0}}  \label{34}
\end{equation}
describe the quadrupole moments of the electric charge distribution in the
elementary cell. The generalization to three dimensions is straightforward,
with the same qualitative properties: The force exerted by some elementary
cell on a charge element $q^{\prime }$ always decreases at least
proportional to $r^{\prime -4}$ with the distance between the charge and the
cell. If the quadrupole moments vanish, as for cubic lattices, the decrease
is even faster by two powers of $r^{\prime }$.This implies that for a given
charge element $q^{\prime }$ the sum over all elementary cells always
converges in $D=1,2,$ or $3$ dimensions. Therefore the force on any ion in
the body is bounded independently from the size of the body, 
\begin{equation}
\left| \vec{F}_{q\prime }\right| _{\max }\leq \sum_{i}\left| F_{q\prime
,V_{i}}^{y}\left( \vec{r}^{\prime }-\vec{r}_{i}\right) \right| \leq
const\sum_{i}\frac{1}{\left| \vec{r}^{\prime }-\vec{r}_{i}\right| ^{4}}.
\label{35}
\end{equation}
Further, the reflection symmetry of the lattice implies that all forces from
cells in a volume that possesses reflection symmetry around the charge
element add up to zero. Hence also in the general case the charge element
feels a force that depends on its distance $R$ to the surface of the body. A
crude estimate gives 
\begin{equation}
\left| \vec{F}_{q\prime }\right| _{R}\leq \sum_{i,\left| \vec{r}\prime -\vec{%
r}_{i}\right| >R}\left| F_{q\prime ,V_{i}}^{y}\left( \vec{r}^{\prime }-\vec{r%
}_{i}\right) \right| \leq const\sum_{\left| \vec{r}\prime -\vec{r}%
_{i}\right| >R}\frac{1}{\left| \vec{r}^{\prime }-\vec{r}_{i}\right| ^{4}}%
\sim R^{D-4}.  \label{36}
\end{equation}
Depending on the dimension, we observe that the force decreases with the
distance from the surface, at least with $1/R$ in three dimensions, if the
quadrupole moments $I_{1},I_{2},I_{3}$ all vanish the decrease is even
stronger. Scaling with the lattice constant $a$ leads us to 
\begin{equation}
\left| \vec{F}_{q\prime }\right| _{R}\leq \left| \vec{F}_{q\prime }\right|
_{\max }\left( \frac{a}{R}\right) ^{4-D}.  \label{37}
\end{equation}
Because the sum over the forces from the short range potential decreases
even faster, the total perturbational forces, which are maximal on the
surface, decrease to about $10^{-3}$ of the surface value within 1000 atomic
layer, which is about 1micrometer. When we integrate the forces (\ref{32},%
\ref{33}) over an elementary cell in order to obtain the mean force on the
cell, we again loose two powers due to (\ref{29}) and (\ref{30}), resulting
in mean forces between two cells $V_{i}$ and $V_{j}$ that decrease at best
proportional to $\left| \vec{r}_{j}-\vec{r}_{i}\right| ^{-6}$. Thus we
conclude that the bulk of the material remains, apart from tidal motion,
unaffected by the forces induced by the gravitational wave. This result has
its origin in the nature of the electromagnetic coupling with its charges of
different signs, and the structure of solid bodies with reflection symmetry
of the elementary cells.

\subsection{Gravitationally coupled matter}

Clearly, the result of the preceding section was due to the conditions (\ref
{29},\ref{30}), but (\ref{29}) does not hold for the attractive
gravitational forces. The acceleration of a test mass is given by 
\begin{equation}
\vec{a}_{G}\left( \vec{r}^{\prime }\right) =\int_{V}d^{D}r\,\;\frac{\rho
_{m}(\vec{r})}{G}\;\frac{\left( \vec{r}^{\prime }-\vec{r}\right) }{\left| 
\vec{r}^{\prime }-\vec{r}\right| ^{5}}h_{bc}\;\left( \vec{r}^{\prime }-\vec{r%
}\right) ^{b}\left( \vec{r}^{\prime }-\vec{r}\right) ^{c},  \label{38}
\end{equation}
where $\rho _{m}$ is the mass density of the body. Assuming that the
wavelength of the gravitational wave is large compared to the dimensions of
a homogeneous massive body, we can take $\rho _{m}$ and $h$ to be constant
and are able to evaluate the integral for simple geometries. For a $+$%
-polarized wave propagating in $z$-direction the maximal acceleration on the
surface is calculated to 
\begin{equation}
\left| \vec{a}_{G}\right| _{\max }=\left\{ 
\begin{array}{l}
\rho _{m}\pi r\left( 1-\frac{1}{\sqrt{1+\frac{l^{2}}{r^{2}}}}\right) \text{
for a cylinder of radius }r\text{ and lenght }l \\ 
\rho _{m}\pi \frac{\tan ^{2}\phi }{\left( 1+\tan ^{2}\phi \right) ^{3/2}}l%
\text{ for the tip of a cone of opening angle }\phi \text{ and height }l \\ 
\rho _{m}\pi \frac{8R}{15}\text{ at the surface a sphere of radius }R.
\end{array}
\right.   \label{39}
\end{equation}
Naturally, there exists an angular dependence of the forces of quadrupole
characteristic. Thus the forces grow linearly with the linear dimensions of
the massive body. The maximal acceleration on the equator of a sphere  is
given  by 
\begin{equation}
\left| \vec{a}_{G}\right| _{\max }^{TT}=\frac{2}{5}\left| h\right| g
\label{40}
\end{equation}
where $g$ is the surface acceleration of the mass. This force will truly
excite the quadrupole modes in the bulk of a body and is able to do work
against the gravitational and electromagnetic forces that keep the body
together. Comparing this with the tidal acceleration in the PFR system, 
\begin{equation}
\left| \vec{a}_{G}\right| _{\max }^{PFR}=\frac{1}{2}\omega ^{2}R\left|
h\right| ,  \label{41}
\end{equation}
we see that (\ref{40}) is several orders of magnitude smaller than (\ref{41}%
). In the PFR system the forces drive primarily the tidal motion, only
accelerations of order (\ref{40}) drive the deviations. 

\section{\protect\bigskip Detector Types}

\subsection{Rotational Detectors}

From the structure of the microscopic quantity $Q_{ab}(\tau )$ we can
immediately identify the different types of detectors. If the body is
rotating relative to the TT frame with some frequency $\omega $, then the
momentum contribution 
\begin{equation}
\sum_{s}\frac{1}{2m_{s}}p_{a}^{(s)}p_{b}^{(s)}\sim \omega ^{2}\cos 2\omega t
\label{23}
\end{equation}
dominates and gives rise to a response of order $h$. This type of detector
was first suggested by Braginsky \cite{braginsky,mtw}. We estimate the
response of a rotating mass to the gravitational wave using our model. From (%
\ref{18}) we obtain a momentum input to the center of mass of 
\begin{equation}
dP_{z}/dt=h_{0}\cos \left( \omega _{0}t+\varphi \right) \cos 2\omega t\frac{%
\omega ^{2}\omega _{0}l^{2}M}{24c}  \label{23a}
\end{equation}
where $l$ is the length of the bar, $M$ its mass, and $h_{0},$ $\omega _{0},$
and $\varphi $ are the amplitude, frequency and phase difference of the
gravitational wave, respectively. This is for the case where the wave vector
is perpendicular to the plane of rotation. In the ideal case $\omega \approx
\omega _{0}/2,\varphi =0$ or $\pi $, the mean energy input is given by 
\begin{equation}
\dot{E}=\pm h_{0}\frac{\omega _{0}^{3}l^{2}M}{192}.
\end{equation}
Note that the change can be of either sign, thus the gravitational wave
cannot only be absorbed, but can also stimulate the emission of
gravitational waves from the system. For reasonable values (bar of $1m$, $%
1-100kg$ mass, $\omega /2\pi \sim 10-1000$ Hertz, $h_{0}\sim 10^{-20}$) the
attainable energy input ranges in about 
\begin{equation}
\dot{E}\sim 10^{-13}...10^{-9}W.
\end{equation}
Though this is quite large compared to the resonant detector, it seems
questionable whether this change in the rotational energy of the bar can be
measured. Certainly the acceleration of the rotating bar in the direction of
the wave, given by (\ref{23a}), is too small to be detectable. This gives
additional justification to the negligence of the acceleration of the center
of mass in the transformation to the PFR system \cite{mtw}.

\subsection{Resonant Detectors}

For a non-rotating mass, the lowest order contribution stems from the time
derivatives of the positions and momenta induced by the wave and thus leads
to a response of order $h^{2}$. This is the Weber detector. Our
considerations have shown that the energy input into such a detector occurs
primarily at its surface. This implies that when resonance is discussed, we
have to take the time into account that is needed to transport form the
surface to the interior of the body. For a material with high Q-factor the
velocity of transport is given by the velocity of sound. Thus it takes a time
\begin{equation}
T_{v}=\frac{L}{2v_{s}}
\end{equation}
where $L$ is the diameter of the body and $v_{s}$ the velocity of sound
before energy reaches the center of the body. This time limits the onset of
resonance. For example, for GRAIL \cite{grail} with diameter $L=3m$, $%
v_{s}\approx 4000m/s$ we have $T_{v}\approx 3/8$ milliseconds. Thus this
time scale will play a role for the detection of millisecond pulses. In
order to arrive at the resonant amplitude, pulses must be considerably
longer than $T_{v}$. How many oscillations are needed to arrive at the
maximal resonant amplitude, can also be estimated in the normal-mode model. 

In general, if a weakly damped oscillator characterized by a frequency $%
\omega _{0}$ is excited by a force $F=\varepsilon \omega ^{2}sin\left(
\omega t\right) $ near resonance, $\omega \approx \omega _{0}$, then if the
oscillator is initially at rest, the subsequent maxima of the amplitude of
the oscillator will at the beginning follow a linear law 
\begin{equation}
\left| A_{max}\right| \sim \frac{\varepsilon \omega ^{2}}{2\omega _{0}}t
\end{equation}
until the resonant amplitude of order $C=\varepsilon C_{0}$ is reached after
a transient time 
\begin{equation}
T_{trans}\approx \frac{2C_{0}}{\omega }\approx \frac{2C_{0}}{\omega _{0}}.
\end{equation}
Thus e.g. it takes for a signal of frequency $\omega =1000Hz$ about $2s$
until it is enhanced by a factor $C_{0}=1000$. Thus $2000$ oscillations are
needed before the maximal resonant level is reached. In general the
effective cross section of a resonant mass detector is proportional to $%
\left( C_{0}\right) ^{2}$, but for pulses shorter than $T_{trans}$ an
additional factor $\left( T/T_{trans}\right) ^{2}$ must be taken into
account. This is relevant to the detection of millisecond pulses with a
detector like GRAIL that operates at a fundamental frequency of $650Hz$ . 

A further point to consider certainly is the impurity of the material. When
grains of material stick together, we also have to consider internal
boundary that behave like surfaces. But because this results in random
effects, the impurity will hardly improve the behavior of the detector.

\subsection{Microscopic Detector}

We have shown that in the TT system the metric effects on the
electromagnetic coupling cancel in the bulk of a massive body, in the PFR
system the deviations from the tidal motion are driven only at the surface.
This raises the question whether this type of surface effect could be used
in some other way for detection of gravitational waves. Similar to
expression (\ref{40}) for the gravitationally coupled matter, we can
estimate the acceleration induce by the gravitational wave on the surface of
an ion lattice by 
\begin{equation}
\left| \vec{a}_{EM}\right| _{\max }^{TT}\sim \left| \frac{Z_{+}Z_{-}}{m_{\pm
}d^{2}\pi \varepsilon _{0}}h\right|   \label{42}
\end{equation}
where $Z_{+},Z_{-}$ and $m_{+},m_{-}$ are the charges and the masses of the
ions, respectively, and $d$ the lattice constant. A geometric factor of
order $1$ has to be included in addition. This factor will depend on the
structure of the lattice and the precise behavior of the repulsive
potential. The numerical value of (\ref{42}) is of the order of $h\cdot
10^{16}ms^{-2}$ (for KCl) which is several orders of magnitude larger than (%
\ref{41}). This suggests that the piezoelectric effect might be used for
experimental verification, in a similar way like a pressure sensor works.
For $h$ in the order of $10^{-22}$ accelerations are of order $10^{-6}ms^{-2}
$ and possibly are within the reach of sensitive detectors. If we consider a
piezo crystal under strain our analysis has to be revised. Because we then
have a dipole moment in each elementary cell, the leading order of the
forces will be proportional to $r^{\prime -3}$ as compared to $r^{\prime -4}$
in (\ref{32}) and (\ref{33}). Due to the uniform direction of the dipoles
and the broken refection symmetry the contributions in the integration will
no longer cancel, so we expect that the forces grow logarithmically with the
dimension of the crystal and the excitation occurs truly throughout the
bulk. We hope to be able present a detailed analysis of this case, together
with possible detector design, soon elsewhere.

\section{Conclusions}

We analyzed the resonant mass gravitational wave detector form a microscopic
point of view, using the wave guide (TT) frame of reference along with the
conventional PFR system. In the TT system the variation of the Coulomb field
of the constituent charges of the body gives the dominant contribution, the
resulting forces agrees with those driving the deviation from the tidal
motion in the PFR system.  For an electromagnetically coupled body with
reflection symmetry,  this force distribution is such that the forces are
restricted to a small surface layer, with no bulk force. Hence the relevant
energy input occurs at the surface only. This effect has its origin in the
fact that the pattern of the tidal forces comes close to a null mode of the
system, not doing work against the coupling forces, as we saw from a linear
chain model. This result is not in contradiction to the standard normal mode
analysis, rather it reflects a local property of the response compared to
global nature of normal modes.  For the gravitationally coupled body, we
observe a linear force law and bulk excitation, but the part that causes
resonant  response \ is orders of magnitude smaller than expected from the
tidal forces in the PFR system. 

Regarding resonant mass detectors, the local nature of the driving force
makes it necessary to consider the time scale on which the energy is
transported into the interior. This time scale is set by the velocity of
sound. As a result, it might take too long a time for a resonant detector to
be excited measurably by short\ millisecond pulses as are  emitted by
collapsing stellar objects. 

Finally, we presented ideas for a new type of microscopic detector that
employs the piezo effect. The force pattern on a crystal is basically that
of  a quadrupole-like distributed pressure change, the surface force is
independent of the size of the crystal. For a crystal under strain the 
force even acts throughout the bulk. An analysis how the desired sensitivity
can  be achieved was outside the scope of the work presented here.

\end{document}